# Bayesian parameter inference and uncertainty quantification for a computational pulmonary hemodynamics model using Gaussian processes

Amirreza Kachabi [a], Sofia Altieri Correa [a], Naomi C. Chesler [a], Mitchel J. Colebank [a,b,*]

[a] Edwards Lifesciences Foundation Cardiovascular Innovation and Research Center, Department of Biomedical Engineering, University of California, Irvine, CA, USA

[b] Department of Mathematics, University of South Carolina, Columbia, SC, USA

ABSTRACT

Subject-specific modeling is a powerful tool in cardiovascular research, providing insights beyond the reach of current clinical diagnostics. Limitations in available clinical data require the incorporation of uncertainty into models to improve guidance for personalized treatments. However, for clinical relevance, such modeling must be computationally efficient. In this study, we used a one-dimensional (1D) fluid dynamics model informed by experimental data from a dog model of chronic thromboembolic pulmonary hypertension (CTEPH), incorporating measurements from multiple subjects under both baseline and CTEPH conditions. Surgical intervention can alleviate CTEPH, yet patients with microvascular disease (e.g., remodeling and narrowing of small vessels) often exhibit persistent pulmonary hypertension, highlighting the importance of assessing microvascular disease severity. Thus, each lung was modeled separately to account for the heterogeneous nature of CTEPH, allowing us to explore lung-specific microvascular narrowing and resistance. We compared inferred parameters between baseline and CTEPH and examined their correlation with clinical markers of disease severity. To accelerate model calibration, we employed Gaussian process (GP) emulators, enabling the estimation of microvascular parameters and their uncertainties within a clinically feasible timeframe. Our results demonstrated that CTEPH leads to heterogeneous microvascular adaptation, reflected in distinct parameter shifts. Notably, the changes in model parameters strongly correlated with disease severity, especially in the lung previously reported to have more advanced disease. This framework provides a rapid, uncertainty-aware method for evaluating microvascular dysfunction in CTEPH and may support more targeted treatment strategies within a timeframe suitable for clinical application.

## 1. Introduction

Computational simulations have emerged as valuable tools in the medical field, offering critical insights that are impractical or infeasible to obtain through clinical measurements. These advanced models can be tailored to individual patients and may lead to personalized treatments based on their predictions. However, before making decisions based on these model predictions, the models must be calibrated using patient-specific data. Model calibration, also referred to as an inverse problem or parameter estimation, involves assessing the consistency between patient-derived measurements and the predictions generated by the model. Formal inference requires an assumption about the statistical nature of the parameters. Different studies in the biosciences have attempted to calibrate their models using either a frequentist, classical statistics approach ([1]; [2]; [3]) or a Bayesian approach ([4]; [5] [6]). While both methods provide mechanisms for uncertainty quantification (UQ), Bayesian methods offer full posterior distributions that support more informative and clinically relevant interpretation of uncertainty compared to the fixed-point estimates and confidence intervals used in frequentist approaches [7]. These models often involve numerically solving partial differential equations (PDEs) to calculate the posterior density for UQ, which can be computationally intensive, sometimes taking several months [8,9]. This makes it challenging to implement within a reasonable clinical timeframe, which would be on the order of minutes or a few hours.

One way to accelerate PDE solutions and solve the inverse problem is using statistical emulation ([10]; [11]). Emulation, which approximates the expensive simulator with a computationally cheaper surrogate model, offers an efficient approach to approximating complex systems, significantly lowering the cost and time of simulations and enables the solution of large-scale or high-dimensional problems. Statical emulations are particularly useful for real-time applications and provide valuable insights into system behavior while ensuring accurate approximation of inverse solutions. Several studies used different techniques to build emulators, such as neural networks (NNs) [12], Gaussian processes (GPs) [13] and polynomial chaos expansions (PCEs) [14]. While PCEs are better suited for low-dimensional problems [15], GPs excel in high-dimensional spaces and perform well with small training datasets, whereas NNs require large datasets for training [16,17].

* Corresponding author. Department of Mathematics, University of South Carolina, Columbia, SC, USA.
*E-mail address:* mjcolebank@sc.edu (M.J. Colebank).

Complex hemodynamic models are particularly useful in understanding biomechanically complex diseases, such as pulmonary hypertension (PH). This condition is characterized by an elevated mean blood pressure (exceeding 20 mmHg) in the main pulmonary artery (MPA) at rest. It is commonly associated with vascular remodeling of both large and small pulmonary arteries. Among the various subgroups of PH, chronic thromboembolic pulmonary hypertension (CTEPH) is a notable form that occurs when blood clots (thromboembolisms) obstruct the pulmonary arteries, leading to increased pulmonary blood pressure. While most cases of CTEPH can be effectively treated with surgery, persistent PH can occur in patients suffering from additional microvascular disease [18]. The severity of small-vessel arteriopathy has been proposed to have a significant impact on the persistence of PH and survival following surgery [19]. However, the current standard preoperative assessment of CTEPH cannot detect the presence of small-vessel disease reliably, nor does it accurately predict the postoperative outcome. Therefore, there is an unmet need for assessing the degree and presence of small-vessel disease in CTEPH, inviting new computational approaches to elucidating this mechanism.

Thus, the goal of this study is to use computational hemodynamic modeling to investigate microvascular disease in CTEPH and examine the impact of CTEPH on microvasculature model parameters. To achieve this, we first construct an emulator using GPs for a one-dimensional (1D) fluid dynamics model of the pulmonary circulation, trained on previously published dog data. For each subject, asymmetric, binary-structured trees are appended to the terminal large vessels to represent the microvasculature. Using the trained GP emulators, we calibrate the model to measured data within a Bayesian framework and identify potential correlations between the inferred parameters and disease severity. We also account for uncertainty in the simulated outputs by propagating posterior distributions, enabling uncertainty-aware parameter estimation and prediction.

To our knowledge, this is the first study to apply GP-based Bayesian inference to investigate microvascular disease in a computational model of CTEPH. Unlike previous studies that apply a single set of parameters to the entire pulmonary vasculature, we independently modeled the left and right lung vasculature to capture the heterogeneous nature of CTEPH. Furthermore, our study is the first to train a GP model within a reduced multi-output space using principal component analysis (PCA) in the context of cardiovascular fluid dynamics. Overall, this work provides a rapid robust framework for integrating limited, noisy data with a complex hemodynamic model. By combining state-of-the-art statistical emulation, dimensionality reduction, and Bayesian inference, we offer new insights into subject-specific parameters indicative of geometric adaptation in the microvasculature due to CTEPH.

## 2. Methods

### 2.1. Data collection

We use animal data originally collected and reported in Mulchrone et al. [20]. In brief, five male dogs were subjected to repeated microsphere injections into the pulmonary circulation to imitate CTEPH development. Additional details can be found in Mulchrone et al. [20]. The data used from the animal study include magnetic resonance angiography (MRA) data, systolic and diastolic main pulmonary artery (MPA) pressure ($p_{sys}^{MPA}, p_{dia}^{MPA}$), time series flow in the MPA, left pulmonary (LPA), and right pulmonary (RPA) arteries ($\boldsymbol{q}^{MPA}, \boldsymbol{q}^{LPA}, \boldsymbol{q}^{RPA}$), and time series MPA area ($\boldsymbol{A}^{MPA}$) in both pre-CTEPH (baseline) and post-CTEPH (CTEPH) stages. We use the baseline MRA data to reconstruct a computational domain for each animal, similar to our previous study [1]. The computational domain includes a connected network of large arteries, which are described by a connectivity matrix and vessel dimensions (i.e., length and radius). We note that, in contrast to this prior study, we use the baseline geometry for all simulations, thus isolating the differences between baseline and CTEPH results (e.g., differences in material properties) from changes in large vessel dimensions or connectivity obtained on sequential MR imaging. The Supplementary Table S1 summarizes the total number of vessels and the number of terminal vessels for each subject. Due to differences in image resolution among subjects, the total and terminal vessel counts vary between them. Additionally, the pulmonary vasculature 1D centerlines for all five subjects are shown in the Supplementary Fig. S1. The microcirculation, which is infeasible to identify by MR imaging, is described using the structured tree model, described later [21].

### 2.2. Data processing

We ensured that the average measured flow was conserved by shifting the measured MPA flow to match the averaged RPA and LPA flows. We also adjusted the minimum flow in certain animals to ensure that the input flow started close to zero for all cases, as there were no reported cases of pulmonary valve regurgitation. All measured MPA flows are shown in the Supplementary Fig. S2. We denote subjects (dogs) 1–5 as D1-D5 respectively. D2 and D4 had clear measurement errors; we rebuilt these waveforms while keeping the maximum and minimum MPA area values the same. Lastly, for D3 at baseline, there was an abnormally long recorded cardiac cycle of 1.4(s), where values from 0.8 to 1.4(s) were constant zero flow. Therefore, we truncated the cardiac cycle to 0.8(s), consistent with the cardiac cycles of other animals.

### 2.3. Governing equations and boundary conditions

We simulated pressure, flow, and vessel area in the vasculature captured by MR by solving the 1D Navier-Stokes equations. We assume that all blood vessels are straight, cylindrical, and impermeable, and that blood flow through the arterial network is axisymmetric, incompressible, Newtonian, and laminar. We then have the governing equations

$$\frac{\partial Q}{\partial x} + \frac{\partial A}{\partial t} = 0, \tag{1}$$

$$\frac{\partial Q}{\partial t} + \frac{(\gamma+2)}{(\gamma+1)} \frac{\partial}{\partial x}\left(\frac{Q^2}{A}\right) + \frac{A}{\rho}\frac{\partial P}{\partial x} = -\frac{2\pi\mu(\gamma+2)}{\rho}\frac{Q}{A}, \tag{2}$$

where $x$ (cm) and $t$ (s) represent the axial and temporal coordinates, and $P(x,t)$ is the transmural blood pressure (g/cm s$^2$) [22]. $Q(x,t)$ is the volumetric flow (mL/s) and $A(x,t)$ is the cross-sectional area (cm$^2$). The parameters $\rho$ and $\mu$ are blood density and dynamic viscosity that are set 1.03 (g/mL) and 0.03 (g/cm s), respectively [23]. Furthermore, we assumed the shape of the blood flow velocity follows a power-law axial profile. We assume $\gamma = 5$ as described in our previous work [1]. We assume a linear stress-strain relationship to close the system of equations, where the vessel stiffness, $K$ (g/cm s$^2$) was analytically obtained by using the measured systolic pressure and area in the pressure-area relationship as written in equation (3):

$$P(x,t) = K\left(\sqrt{\frac{A(x,t)}{A_{dia}}} - 1\right) + P_{dia}. \tag{3}$$

where $A_{dia} = \pi r_{dia}^2$ (cm$^2$) is the lumen area obtained from MR at the diastolic pressure $P_{dia}$ (mmHg). $K$ was fixed throughout the entire vasculature (both large and small arteries), as described in our previous work [1].

The PDE system in equations (1) and (2) requires boundary conditions at each vessel inlet and outlet, including at vessel junctions and terminal branches at the end of the MR imaging domain. For the MPA inlet, we use a subject-specific measured time series flow as a boundary condition. At each vessel junction, we enforce flow conservation and pressure continuity [1]. At the outlet of the terminal large arteries, we impose a synthetic vascular tree representative of the microvasculature using the structured tree model [21].

The structured tree attached to each terminal vessel is a construction of an asymmetric binary tree, where at each bifurcation the radii of the two offspring vessels are scaled by factors $\alpha$ and $(0<\beta<\alpha<1)$, respectively (Fig. 1). The network is truncated at a terminal radius, $r_{min}$, set to be 0.005 (cm) [24]. The ST network can be formulated using the area ratio ($\zeta$) and Murray exponent ($\eta$) [21]

$$\zeta = \frac{A_{d2}}{A_{d1}}, r_p{}^{\eta} = r_{d1}{}^{\eta} + r_{d1}{}^{\eta}. \tag{4a,b}$$

The subscripts $p$, $d1$, and $d2$ represent the parent, larger offspring, and smaller offspring branch in the tree, respectively, and the offspring areas, $A_{d2}$ and $A_{d1}$, correspond to the smaller and larger offspring vessels in each bifurcation. The area ratio $\zeta$ was determined using the MR imaging data and was calculated as the median value among all bifurcations across the large artery tree in all subjects, resulting in $\zeta = 0.6$. We note that $\eta$ and $\zeta$ are related to $\alpha$ and $\beta$ via the relationship [21]

$$\alpha = \left(1+\zeta^{\eta/2}\right)^{\frac{-1}{\eta}}, \beta = \alpha\sqrt{\zeta} \quad . \tag{5}$$

hence $\alpha$ and $\beta$ are functions of $\eta$ and $\zeta$. For each vessel $i$ in the ST, the reference radius is calculated by

$$r_i = r_{term} \cdot \alpha^g \cdot \beta^h \quad . \tag{6}$$

where $r_{term}$ is the radius of the terminal vessel to which the structured tree is attached, $g$ is the degree of generation in the $\alpha$ path and $h$ is the degree of generation in the $\beta$ path. Also, for each given vessel $i$, the length of the vessel, $l_i$, is assumed to vary linearly with the vessel radius and is calculated by using the length-to-radius ratio ($lrr$):

$$l_i = r_i \cdot lrr \quad . \tag{7}$$

Briefly, the fluid mechanics in microvasculature are assumed to be viscous dominant, leading to a linear momentum balance equation without an inertial term. This viscous-dominant system is transformed into the frequency domain, with angular frequencies $\omega_j = 2\pi j/T$ (radian/s) where $T$ (s) is the length of the cardiac cycle and $j$ is an integer. The small vessels are also assumed to be linearly elastic, with the same stiffness as the large vessels (equation (3)). The solution to the linear system of periodic PDEs constitute a wave equation, with solutions represented by a combination of sines and cosines [24]. The structured tree provides an impedance boundary condition, $Z_{tot}(\omega) = P(\omega)/Q(\omega)$, to the 1D system by using the analytical wave equation solutions of pressure, $P(\omega)$, and flow, $Q(\omega)$, in combination with continuity of pressure and flow conservation at each bifurcation. To calculate $Z_{tot}(\omega)$, the structured tree geometry is generated up until a radius $r_{min}$ and the radius, length, and stiffness of each vessel branch is prescribed. Then, the impedance of the smallest branches of the structured tree are calculated via

$$\overline{Z}(0) = \frac{8\mu(r_0)lrr}{\pi r_0{}^3}. \tag{8}$$

where $\overline{Z}(0)$ is the mean (zero-frequency) impedance, $\mu(r_0)$ is the radius dependent viscosity, and $r_0$ is the reference radius [25]. The impedances, similar to resistances, are combined in series and in parallel using circuit theory analogues, resulting in a total impedance of each ST, $Z_{tot}(\omega)$, that is used as a boundary condition for the PDE system. Finally, the frequency domain impedance is coupled to the time domain PDEs using a convolution interval (see Olufsen [21] for details).

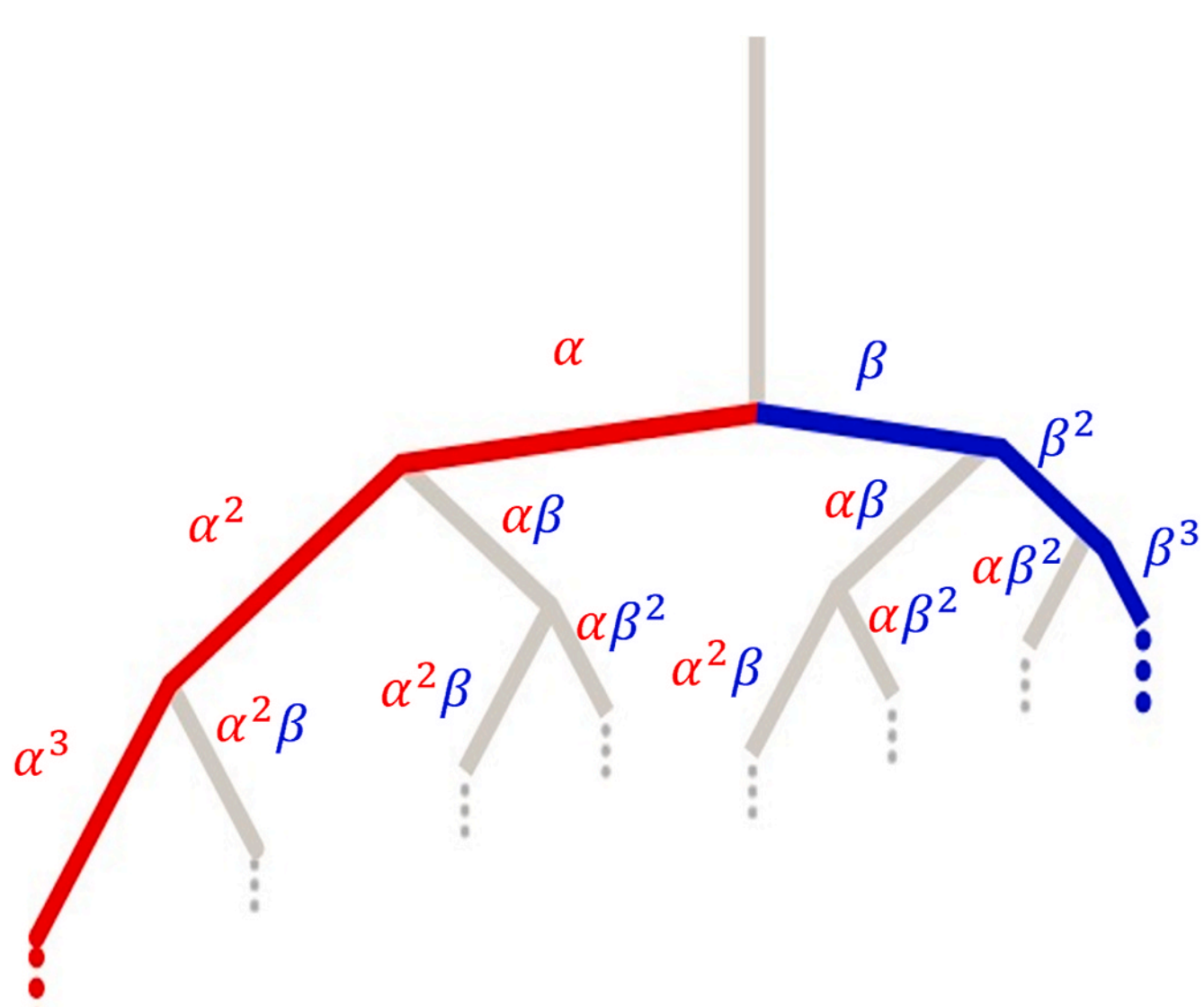

**Fig. 1.** Schematic of the structured tree boundary condition.

### 2.4. Model parameters

The model includes two different sets of parameters. The constant fluid parameters are fixed based on literature, such as $\mu$, $\rho$ and $r_{min}$ or obtained from data, like $\zeta$, $K$, and $T$, for each individual subject. These fixed parameters and their corresponding values for each subject under both baseline and CTEPH conditions are presented in Table 1. A different set of parameters describing the microvasculature are inferred to match each subject's hemodynamic data, including the Murray exponent ($\eta$), and length to radius ratio ($lrr$). Since CTEPH is a heterogeneous disease that can affect each side of the lung differently, we used independent $\eta$ and $lrr$ parameters for the left and right lungs. This also parallels findings in Mulchrone et al., which identified an unequal number of microspheres in the left and right lung. In total, our parameter set to be inferred is four-dimensional, comprised of $\boldsymbol{\theta} = \{\eta_L, \eta_R, lrr_L, lrr_R\}$, where $L$ and $R$ denote the left and right lung parameters.

### 2.5. Emulation

We overcome the computational burden of using our expensive PDE simulator by using GP emulation. GPs are beneficial in comparison to other emulators because they are more robust at capturing model dynamics using smaller datasets for training [26]. Briefly, a stochastic process, $f = f(\boldsymbol{\theta})$, is defined as a GP if the finite collection of random variables $(f(\boldsymbol{\theta}_1), \ldots, f(\boldsymbol{\theta}_n))$ are jointly normal for inputs $\boldsymbol{\theta}_i \in \mathbb{R}^p$, where $i = 1, \ldots, n$ are the realizations and $p$ is the parameter dimensionality. We write the GP as $f \sim GP(\boldsymbol{m}, \boldsymbol{K})$, where $\boldsymbol{m}$ is the mean function and $K = \left[k(\theta_i, \theta_j)\right]_{i,j=1}^{n}$ is the $n \times n$ variance-covariance matrix of $f$ based on the kernel function $k(\theta_i, \theta_j)$. In GP models, the input $\boldsymbol{\Theta}$ an ($n \times d$ matrix) are mapped to outputs $\boldsymbol{y} = (y_1, \ldots, y_n)$ (an $n$-vector) through latent noiseless functions $\boldsymbol{f}$ [26–28].

We trained a GP model on 3,000 datasets from the PDE simulator with different parameter combinations, $\boldsymbol{\theta}_{Sampling} = \{\eta_L, lrr_L, \eta_R, lrr_R\}$, using a Latin hypercube design. Table 2 presents the upper and lower bounds selected for the datasets. To obtain the intervals for sampling, we combined calculations from the large vessel imaging data and previous literature values [24,29,30]. For the Murray exponent ($\eta$) value, we solved equation (4b) for $\eta$ at each bifurcation for each subject using a Newton-Raphson minimization in MATLAB (Mathworks, Nantick MA). By aggregating the $\eta$ values across various bifurcations and subjects, we initially obtained a range of [0.5,4]. We narrowed this interval to [1.5,3] based on physiological considerations and the operational limits of our 1D fluid model [21,31]. For the parameter *lrr*, the range we obtained from data in the large vessels was [0.1,44]. Olufsen et al. [24] suggested that *lrr*, values can reach up to 50 for both systemic and pulmonary circulation. To accommodate this and provide a broader interval, we extended the range to [2.5,70].

**Table 1**
Fixed parameter values for all subjects under baseline and CTEPH conditions.

| Parameters | Physiological meaning | D1 | | D2 | | D3 | | D4 | | D5 | |
|---|---|---|---|---|---|---|---|---|---|---|---|
| | | Base | CTEPH | Base | CTEPH | Base | CTEPH | Base | CTEPH | Base | CTEPH |
| $\mu(g/mL)$ | Blood viscosity | 0.03 | 0.03 | 0.03 | 0.03 | 0.03 | 0.03 | 0.03 | 0.03 | 0.03 | 0.03 |
| $\rho(g/cm\ s)$ | Blood density | 1.03 | 1.03 | 1.03 | 1.03 | 1.03 | 1.03 | 1.03 | 1.03 | 1.03 | 1.03 |
| $\zeta$ | Area ratio | 0.6 | 0.6 | 0.6 | 0.6 | 0.6 | 0.6 | 0.6 | 0.6 | 0.6 | 0.6 |
| $r_{min}(cm)$ | Minimum radius | 0.005 | 0.005 | 0.005 | 0.005 | 0.005 | 0.005 | 0.005 | 0.005 | 0.005 | 0.005 |
| $K(g/cm\ s^2)$ | Vessel stiffness | 7.7e4 | 1.01e5 | 9.1e4 | 1.9e5 | 1.08e5 | 1.1e5 | 5.5e4 | 2.03e5 | 4.8e4 | 7.3e4 |
| T $(s)$ | Cardiac cycle | 0.5 | 0.65 | 0.65 | 0.55 | 0.8 | 0.8 | 0.8 | 0.5 | 0.6 | 0.6 |

The simulator outputs include four different time-varying signals: MPA pressure, LPA and RPA flows, and MPA area, making the final output dimension a 35 x 4 matrix, resulting in a total of 140 output features. To reduce computational complexity that would be attributed to multitask learning for our vectorized outputs, we applied PCA to the simulator outputs. The PCA representation of the model outputs is given by

$$y(\boldsymbol{\theta},\boldsymbol{t})=\bar{\boldsymbol{y}}+\sum_{i=1}^{n_{PCA}}c_i(\boldsymbol{\theta})\boldsymbol{\xi}_i(t)+\varepsilon_i(\boldsymbol{t})\ . \tag{9}$$

where $n_{PCA}$ is the number of PCA terms, $\bar{\boldsymbol{y}}$ is the mean of the simulator outputs $\boldsymbol{y}$, $\boldsymbol{\xi_i}$ are the eigenvectors corresponding to the sample covariance matrix of the data, $c_i(\boldsymbol{\theta})$ are the principal component scores, and $\varepsilon_i(\boldsymbol{\theta},\boldsymbol{t})$ represents the error in the PCA approximation. Since the principal components of the outputs are independent, we emulate the PCA scores, $c_i(\boldsymbol{\theta})$ using independent GPs ([27].). We employed min-max scaling for both the parameter inputs and simulator outputs to account for differences in magnitudes and ensure robust emulator building [32]. We used a zero mean GP function, similar to previous work [26]. To apply PCA to the output signals, we treated them as a single combined output, as they were generated from a PDE model and are inherently correlated due to the underlying physics. By applying PCA to the full signal, we transform the coupled PDE outputs into independent components with which we may build our emulator. We performed PCA on each subject's simulator data set individually and fixed the number of principal components based on maximum required components to capture more than 99.9 % of the original variance across all datasets. This resulted in $n_{PCA}=20$ principal components, reducing the original dimensionality down from 140, and thus emulating the PCA data using 20 independent GPs for each $c_i(\boldsymbol{\theta})$ with a Matérn covariance kernel ($\nu=5/2$).

To train the GPs, we minimize the negative log likelihood of the model, calculated using the marginal log likelihood of the observed data, using the Adam optimizer with a learning rate of 0.1. We monitor convergence across 1,000 iterations. All GP implementation and training was done in the GPytorch infrastructure ([33].) and is available at https://github.com/AmirrezaKachabi.

## 2.6. Inverse problem and model calibration

We use the collection of subject-specific measurements, $\boldsymbol{y}=\left\{p_{sys}^{MPA},p_{dia}^{MPA},\boldsymbol{q}^{LPA},\boldsymbol{q}^{RPA},\boldsymbol{\varepsilon}^{MPA}\right\}$, for the inverse problem. Here, $\boldsymbol{\varepsilon}^{MPA}$ is the MPA relative area change obtained from dynamic area using $(t)=\frac{A(t)-A_{dia}}{A_{dia}}$. We use relative area change instead of dynamic area to account for the constant use of the baseline geometry in each subject. Further, $\varepsilon(t)$ is suggested to be a non-invasive measure of proximal arterial stiffening and a predictor of mortality in PH [34]. The relative area change signals were converted to percentage values for the inverse problem.

We use a Bayesian framework for parameter inference. We assume the parameters are random variables, where the prior parameter distributions and likelihood function are used to sample from the approximate posterior distribution conditioned on the observed data. This gives rise to Bayes' formula:

$$P(\boldsymbol{\theta}|\boldsymbol{y})=\frac{P(\boldsymbol{y}|\boldsymbol{\theta})P(\boldsymbol{\theta})}{\int P(\boldsymbol{y}|\boldsymbol{\theta})P(\boldsymbol{\theta})d\boldsymbol{\theta}}\ . \tag{10}$$

Given measured data $\boldsymbol{y}$, $P(\boldsymbol{y}|\boldsymbol{\theta})=\mathscr{L}(\boldsymbol{y}|\boldsymbol{\theta})$ is the likelihood function, and $P(\boldsymbol{\theta})$ is the prior distribution for the parameters. The term $\int P(\boldsymbol{y}|\boldsymbol{\theta})P(\boldsymbol{\theta})d\boldsymbol{\theta}$ is the evidence or the normalization factor, which is approximated here using Markov chain Monte Carlo (MCMC). In place of the true likelihood using our PDE system, we use the GP emulator.

We consider two different prior distributions to investigate its impact on the posterior densities. First, we use Gaussian prior distributions derived from the median and standard deviation of the large vessel parameters across all subjects over the physiological ranges defined in Table 2. Specifically, we assume that the priors are $\eta\sim\mathscr{N}(2.13,0.37)$ and $lrr\sim\mathscr{N}(10.7,8)$. For comparison, we also use a uniform prior over the plausible parameter space defined in Table 2. We prescribe different magnitudes of measurement error to each of the four measurement modalities, giving the likelihood function:

$$\mathscr{L}(\boldsymbol{y}|\boldsymbol{\theta})=\frac{1}{(2\pi)^{N_t/2}}\det\left(\boldsymbol{\Sigma}_y\right)^{-1/2}\exp\left(-\frac{1}{2}(\boldsymbol{y}-\widetilde{\boldsymbol{m}}(\boldsymbol{\theta}))^{\top}\boldsymbol{\Sigma}_{\boldsymbol{y}}^{-1}(\boldsymbol{y}-\widetilde{\boldsymbol{m}}(\boldsymbol{\theta}))\right)\ . \tag{11}$$

The likelihood includes $N_t=107$ (2 +3 x 35) data points accounting for the two pressure values, two flow signals and one strain signal corresponding to the original data $\boldsymbol{y}$. The term $\widetilde{\boldsymbol{m}}(\boldsymbol{\theta})$ represents the inverse PCA transformed GP model predictions, and the measurement covariance $\boldsymbol{\Sigma}_{\boldsymbol{y}}\in\mathbb{R}^{N_t\times N_t}$ is a diagonal matrix containing the measurement error for each data source. The measurement error values in $\boldsymbol{\Sigma}_{\boldsymbol{y}}$ updated according to an inverse-gamma distribution [7]. We sample from the approximate posterior using the delayed rejection adaptive Metropolis (DRAM) algorithm in the pymcmcstat package [35]. We run DRAM for 10,000 iterations, with the first 2,000 iterations treated as the burn-in phase. To determine the convergence of the MCMC chains, we use the Geweke test, which compares the first 10% and last 50% of the chain for significant differences in the mean and computes a p-value from a Z-statistic [7]. We compute both credible and prediction intervals to quantify uncertainty in the model outputs. Credible intervals represent the range within which a parameter or model output is likely to lie with a certain probability, given the observed data. In contrast, prediction intervals indicate the range likely to contain a future observation, accounting for both model uncertainty and measurement error, and are therefore wider. Output uncertainty intervals are constructed by propagating uncertainty through the model using the last 2000 posterior samples. A summary of the entire process can be found in Fig. 2.

**Table 2**
Parameter bounds for sampling.

| Parameter | $\eta_L$ | $lrr_L$ | $\eta_R$ | $lrr_L$ |
|---|---|---|---|---|
| Range | $1.5-3$ | $2.5-70$ | $1.5-3$ | $2.5-70$ |

### 2.7. Statistical analysis

To compare the marginal posteriors between baseline and CTEPH and with different priors, we used the Kolmogorov-Smirnov (KS) test with a p-value $<0.05$ considered significant. We used the Mann-Whitney $U$ test to compare the posterior distributions of microvascular parameters of each subject before and after CTEPH. Finally, we used the Pearson correlation coefficient performed based on relative changes between the model outputs from baseline to CTEPH to quantify relationships between model parameters and predictions. We used the flow split ratio ($q_r$), calculated using the formula below, and relative changes in mean pulmonary arterial pressure (mPAP) as the measures of disease severity to investigate the correlation with microvasculature parameters:

$$q_r = \frac{\overline{\boldsymbol{q}}_m^{LPA}}{\overline{\boldsymbol{q}}_m^{LPA} + \overline{\boldsymbol{q}}_m^{RPA}} . \tag{12}$$

where $\overline{q}_m$ denotes the time-averaged flow derived from the model.

## 3. Results

### 3.1. Emulation accuracy

We reserved 5 % of each subject's PDE simulator data as test data to assess emulator accuracy across different subjects, under both baseline and CTEPH conditions, for all four data sources used to train the GP. We compared subjects across conditions because (i) the geometries differ between subjects, and (ii) the inflow profiles driving the PDE model are distinct between baseline and CTEPH. We calculated the log mean squared error (MSE) for each output quantity (i.e., pressure, flows, and area) in both baseline and CTEPH subjects. Note that each data source has different units and order of magnitude.

As shown in Fig. 3(a), all subjects (D1-D5) show similar accuracy in pressure, except for D1 and D5 in baseline, and D1 in CTEPH, which have a lower median log MSE. The same trend exists for the area predictions, as provided in Fig. 3(b). This behavior can be attributed to the direct relationship between pressure and area. Due to the correlation between the flows for a given simulation, the median log MSE for LPA and RPA flow are similar in magnitude across most cases as shown in Fig. 3(c and d). We note that D1 has the largest flow MSE outliers among the subjects especially at baseline. In general, median MSE values are lower in the LPA than the RPA, reflecting better accuracy of the emulator in the left branch. We note that there is variability among subjects in the accuracy of the emulator across output types. On average, there is reasonable agreement between the GP and PDE test data in both baseline and CTEPH conditions. We also investigated the effects of training size and the number of principal components on two different subjects (D1 at baseline and D2 in CTEPH). The results are shown in the supplementary figures (Fig. S3–S4). Overall, across all four data sources, increasing the training size beyond 2000 samples leads to only marginal improvements in accuracy, as indicated by the small decrease in the log MSE. Similarly, increasing the number of principal components improves model performance, but the gains become negligible between 20 and 25 components, supporting our choice of using 20. Although including more components may slightly enhance accuracy, it also increases computational time.

### 3.2. Model calibration

Figs. 4 and 5 illustrate the four measured data sources, along with calibrated GP predictions at the posterior mean, in baseline and CTEPH respectively. We also show 95 % credible and prediction intervals, derived from the last 2000 simulations, for all subjects. The experimental report by Mulchrone et al. [20] observed an increase in MPA pressure from baseline to CTEPH in all animals, a reduction in average LPA flow across all subjects, a decrease in average RPA flow in Subjects D1 and D4, and an increase in average RPA flow in Subjects D2, D3, and D5. Additionally, a reduction in RAC was reported in all subjects from baseline to CTEPH. In Figs. 4 and 5, to provide a clearer view of how CTEPH impacted each data source for individual subjects, the y-axis limits are fixed for each subject across conditions. The agreement between the data and the GP varies among the subjects. Model predictions show excellent agreement with the measured data in D1 (Fig. 4(a)) and D4 (Fig. 4(c)). For D2 (Fig. 4(b)), LPA and RPA flows align well with the time-series data, while the GP predictions for systolic pressure and maximum strain are notably higher than the measured data. Model predictions in D4 (Fig. 4(d)) agree well with the data also, with only a slight underestimation of maximum LPA flow and systolic pressure. As depicted in Fig. 4(e), in D5 the GP model performs exceptionally well for pressure and flows, with only a slight mismatch in maximum strain, which is covered by the prediction interval.

Similar to baseline, there is variability among CTEPH subjects in the agreement between the data and the GP, with generally less agreement compared to the baseline cases (Fig. 5). Systolic pressures match well for D3 (Fig. 5(c)), while for the other subjects the GP predictions are larger in magnitude than both the systolic and diastolic data. For the LPA flow in D1, the model provided an excellent fit, while for the other subjects there are some mismatches in flow magnitude as shown in Fig. 5(a). However, the GP can emulate flow waveform shapes reasonably well in both systole and diastole. Predictions of RPA flow match well in D1, D3, and D4 (Fig. 5(d)), while there are slight mismatches for D2 and D5 (Fig. 5(e)). For strain, the fits for D3 and D4 agree well with the data,

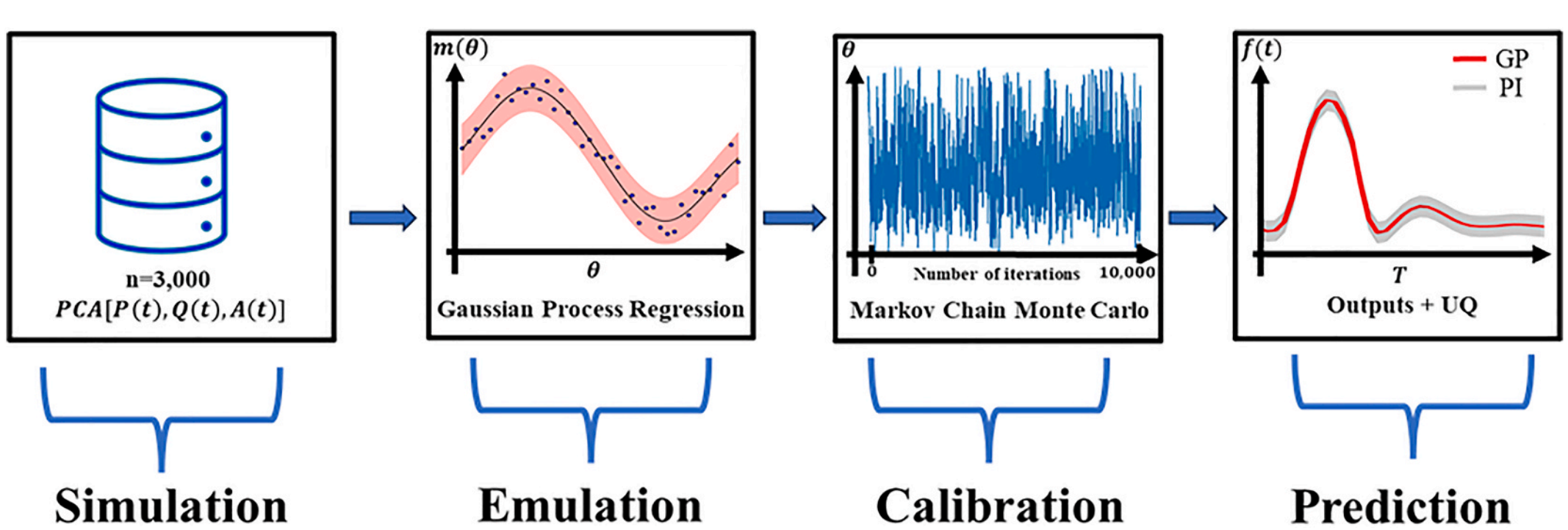

**Fig. 2.** A summary of entire process includes (a) creating a dataset (b) training a GP (forward problem) (c) model calibration (inverse problem) and (d) model predictions with the prediction interval (PI).

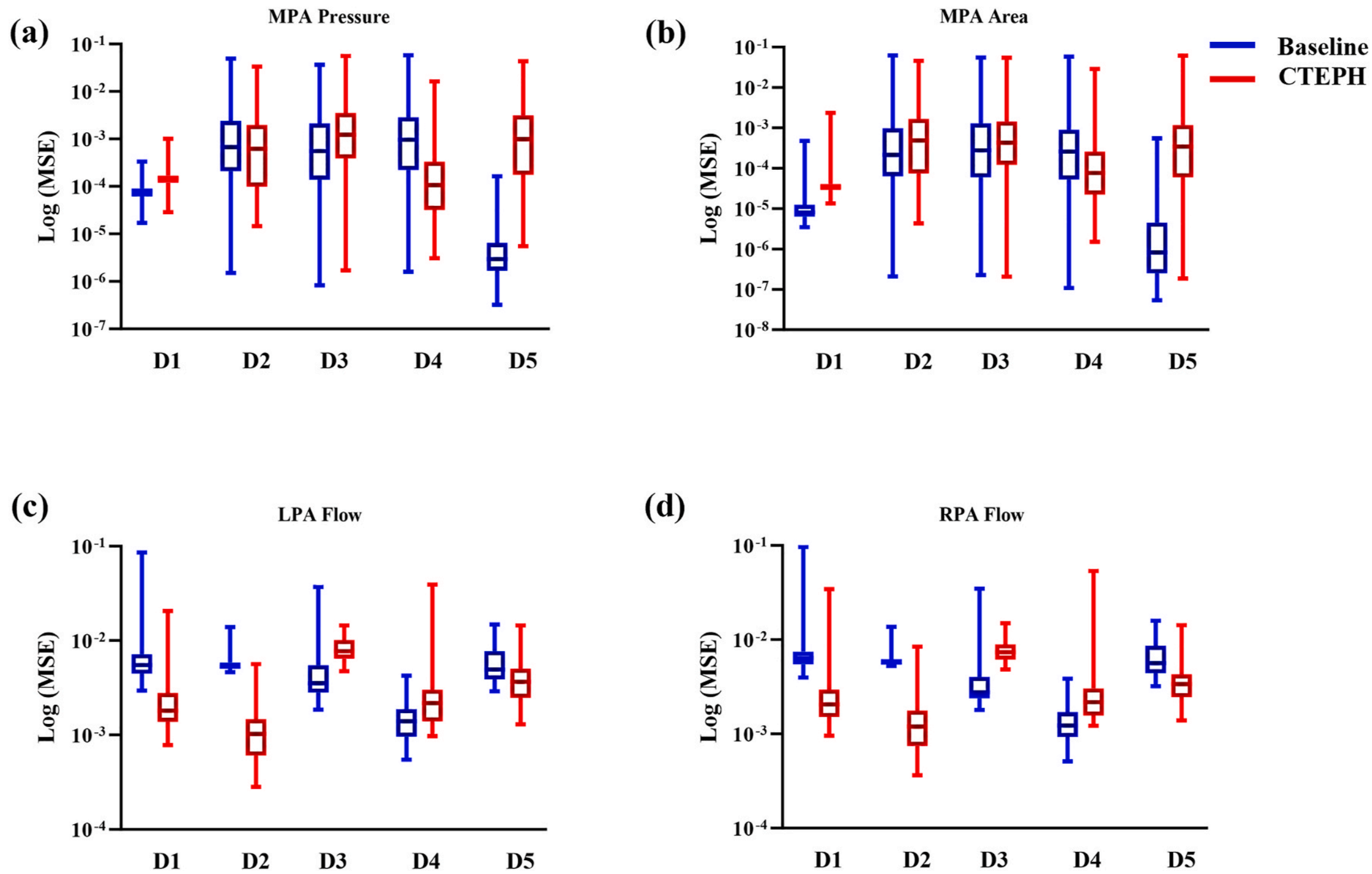

**Fig. 3.** GP emulator Vs. PDE simulator across each output type (pressure, flow, area) and subject (D1 – D5, respectively).

while the GP model overestimated the maximum strain in D1, D2, and D5.

Supplementary Table S2 presents the difference for the static values (systolic and diastolic pressures), as well as the $R^2$ values for the time-series signals (LPA and RPA flows, and MPA strain) between the measured data and the model predictions at the posterior mean. It includes results for each individual subject, along with the absolute average values for all baseline subjects and all CTEPH subjects.

### 3.3. Marginal posterior densities

The marginal posterior distributions shown in Figs. 6–8 have the probability density function (PDF) on the y-axis, representing the probability of each parameter value. The x-axis shows the range of the corresponding parameters and is fixed across all subjects to improve visualization and enable direct comparison. Note that each PDF integrates to unity. We also present supplementary figures (Fig. S5–7) illustrating the maximum a posteriori (MAP) estimates and 90 % credible intervals for each parameter. These additional visualizations help clarify the differences among the posterior distributions. Fig. 6 shows the marginal posterior densities for all subjects in both baseline and CTEPH conditions. We provide the baseline (blue) and CTEPH (red) posteriors for each subject in the same plot for each parameter. The posterior distributions of all model parameters were found to be significantly different (p-value <0.005) between baseline and CTEPH using the KS test. However, Fig. 6 qualitatively suggests that for certain subjects, this difference is not substantial. For example, $\eta_L$ in D1 or $\eta_R$ in D2, D3, and D5, as well as $lrr_R$ in D3 appear similar. Furthermore, the Mann-Whitney $U$ test also supported that CTEPH significantly altered the parameter distributions (p-value <0.005). For the left lung, CTEPH significantly reduced $\eta_L$ across all subjects and increased $lrr_L$ in all subjects except D3. In the right lung, the results were heterogeneous: $\eta_R$ decreased in subjects D1 and D5 but increased in D2, D3, and D4, while $lrr_R$ significantly decreased in D1 and D3 and significantly increased in D2, D4, and D5.

### 3.4. Effect of prior distribution

The results in Fig. 6 were obtained under the assumption of a data-driven, Gaussian prior based on values obtained from MR data. To identify the effects of this assumption, Fig. 7 shows the marginal parameter posteriors using either data driven, Gaussian priors (solid lines) or using uniform priors on the plausible parameter range (dashed lines) in baseline. Results are obtained using the same MCMC routine described earlier. In D1, the posterior densities for both $\eta_L, \eta_R$ were slightly shifted to the left, with minimal impact on either $lrr$. In D2, the prior had no influence on $\eta_L$, $lrr_L$, or $lrr_R$, but it made the posterior for $\eta_R$ more unimodal. D3 was the only case in which the informative prior affected all parameter posteriors at baseline, shifting them to the left. To further examine this, we compared the results using informative and flat priors for D3 and found that, as shown in Supplementary Fig. S8(a), the model predictions based on their respective posterior means were nearly identical. In D4, the informative prior had almost no impact on the posteriors of other parameters but successfully mitigated the long tail observed in the posterior of $\eta_R$. Similarly, in D5, the informative prior did not affect any of the marginal posteriors.

Fig. 8 shows how the prior density impacted these subjects in CTEPH. The Gaussian prior increased the posterior samples of the left lung parameters while decreasing the right lung parameters in D1 and D4. In D2, the shape of the posteriors was mostly affected, with the mode remaining almost unchanged. In D3, the informative prior had no impact on the left lung parameters, but it shifted the right lung parameters to toward smaller values. In D5, all four parameters shifted to the right. These results suggest that inference using the CTEPH data was more sensitive to the choice of prior. Since D1 and D5 showed the most pronounced changes in their posterior distributions between informative and flat priors, we compared the corresponding model predictions for these two subjects in the supplement. As shown in Fig. S8(b), for Subject D1 at CTEPH, the model predictions for pressure and strain were closer to the measured data when using a flat prior, while the flow predictions were similar for both priors. In contrast, as shown in Fig. S8 (c), for Subject D5, the informative prior resulted in more accurate pressure predictions during systole, while the flat prior provided a better fit during diastole. For both flow and strain, predictions using informative and flat priors were nearly identical.

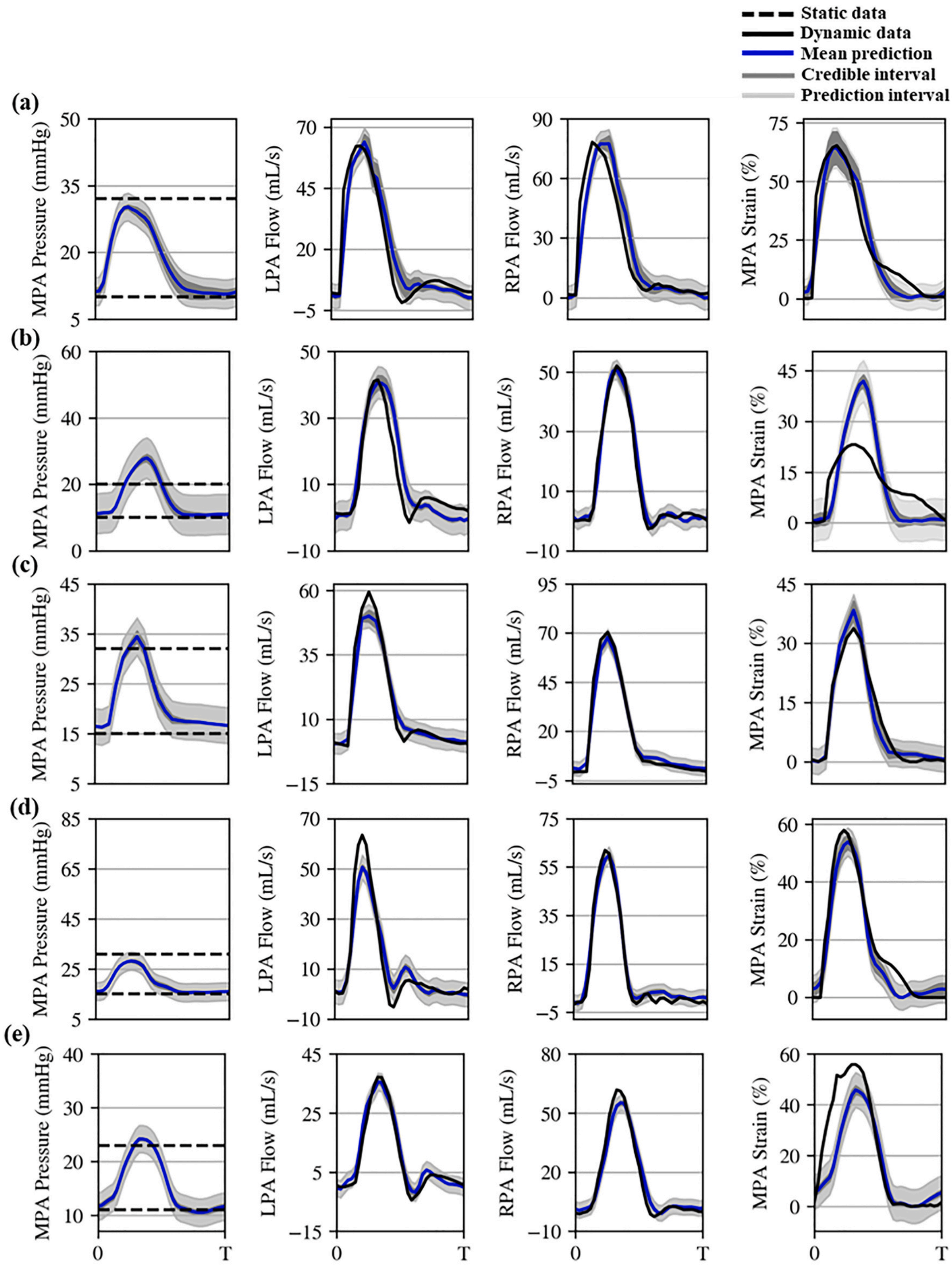

**Fig. 4.** Calibration results for all subjects in baseline. The figure shows the measured static data (dashed lines) and dynamic data (solid lines) across all four sources (black) with dashed lines representing static data and solid lines indicating dynamic data, GP predictions at the posterior mean (blue), the credible interval (dark gray), and the prediction interval (light gray). Subplots are organized as follows: (a) D1, (b) D2, (c) D3, (d) D4, and (e) D5. (For interpretation of the references to colour in this figure legend, the reader is referred to the Web version of this article.)

### 3.5. Parameter correlation

To analyze the microvasculature parameters and their changes with disease severity, we performed Pearson correlation analysis. Correlations values are shown in Fig. 9, with those above 0.80 and below −0.80 highlighted as strong correlations. We found strong positive correlations between $q_r$ (see equation (10)) and left lung parameters ( $\eta_L$ and $lrr_L$) while for the mPAP, we noticed a strong positive relationship just with $lrr_L$. In contrast, the correlations with the $\eta_R$ were relatively weak.

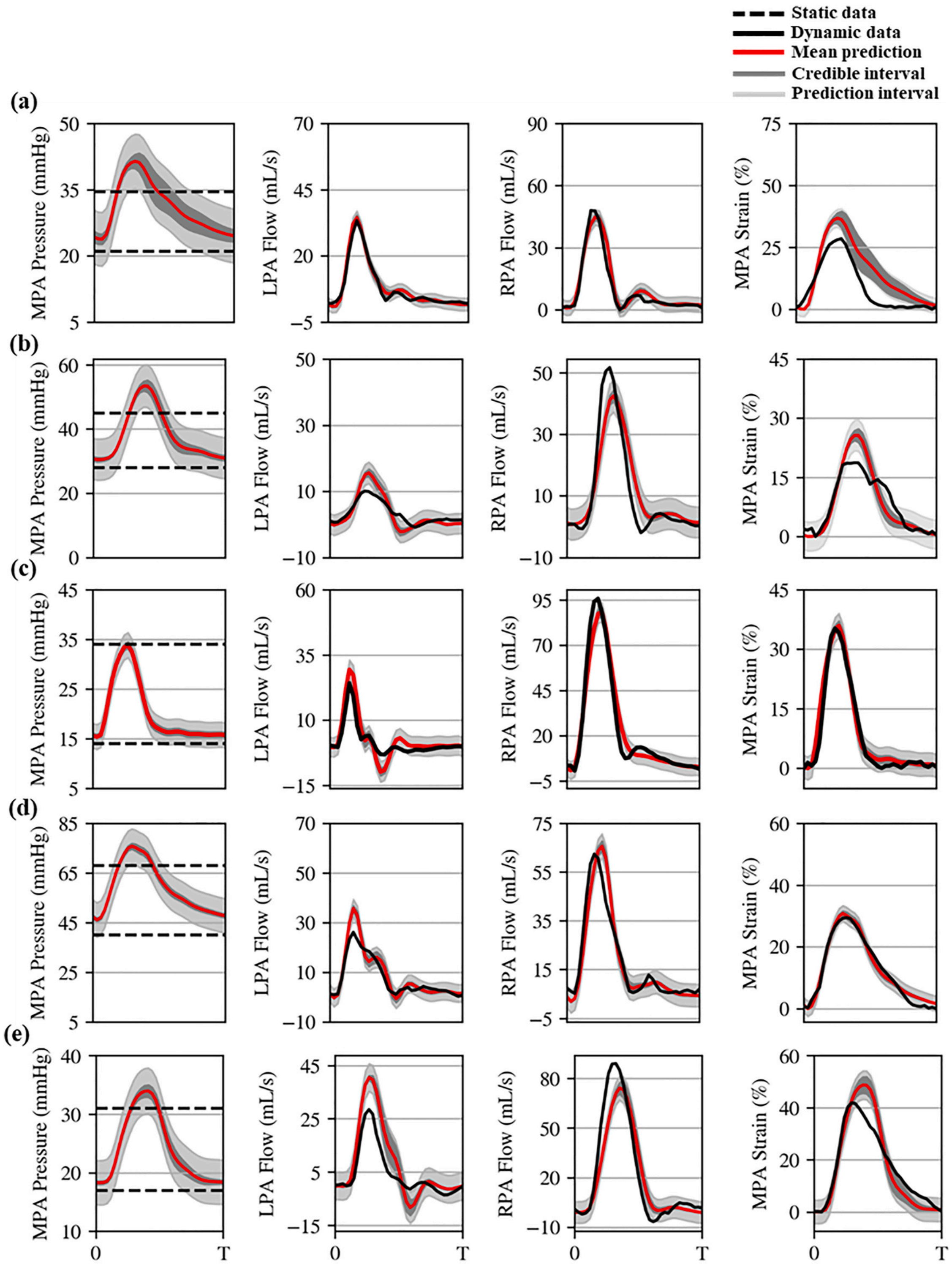

**Fig. 5.** Calibration results for all subjects in CTEPH. The figure shows the measured static data (dashed lines) and dynamic data (solid lines) across all four sources (black) with dashed lines representing static data and solid lines indicating dynamic data, GP predictions at the posterior mean (red), the credible interval (dark gray), and the prediction interval (light gray). Subplots are organized as follows: (a) D1, (b) D2, (c) D3, (d) D4, and (e) D5. (For interpretation of the references to colour in this figure legend, the reader is referred to the Web version of this article.)

Furthermore, to explore clustering between baseline and CTEPH subjects, we generated two-dimensional visualizations using t-SNE, UMAP, and PCA based on ten features: left and right microvasculature parameters, vessel stiffness, systolic and diastolic pressures, average flow at the LPA and RPA, and maximum strain. These visualizations are shown in the supplementary Fig. S9–11. However, due to the small sample size, none of these methods provided a robust separation between the baseline and CTEPH groups. We believe that a larger sample size would significantly improve the reliability and strength of these findings.

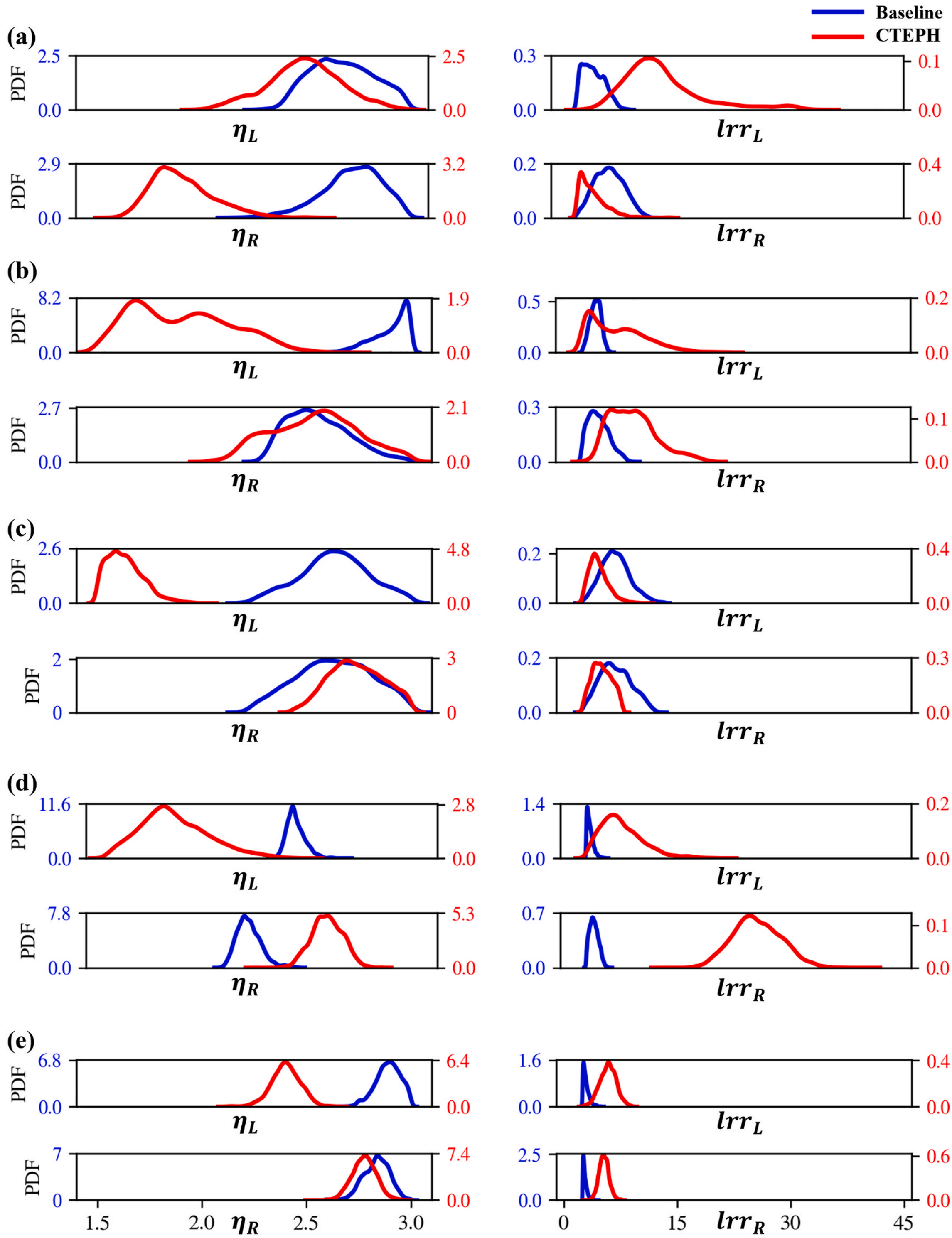

**Fig. 6.** Marginal posterior distributions of the ST parameters for all subjects, shown for baseline (blue) and CTEPH (red). Subplots are organized as follows: (a) D1, (b) D2, (c) D3, (d) D4, and (e) D5. (For interpretation of the references to colour in this figure legend, the reader is referred to the Web version of this article.)

## 4. Discussion

In this study, we emulated a 1D pulmonary circulation model using GPs. Our aim was to computationally investigate the impacts of CTEPH on the microvasculature, a domain that is challenging to study clinically or preclinically using current imaging equipment. We used a lung-specific approach for left and right lung vasculature and, by employing GPs as an emulator for the expensive PDE model, we were able to estimate model parameters describing the microvasculature using Bayesian inference. This approach allowed us to quantify uncertainties in model parameters and emulated outputs, and offers valuable insights about pulmonary vascular narrowing, particularly when working with limited and noisy data.

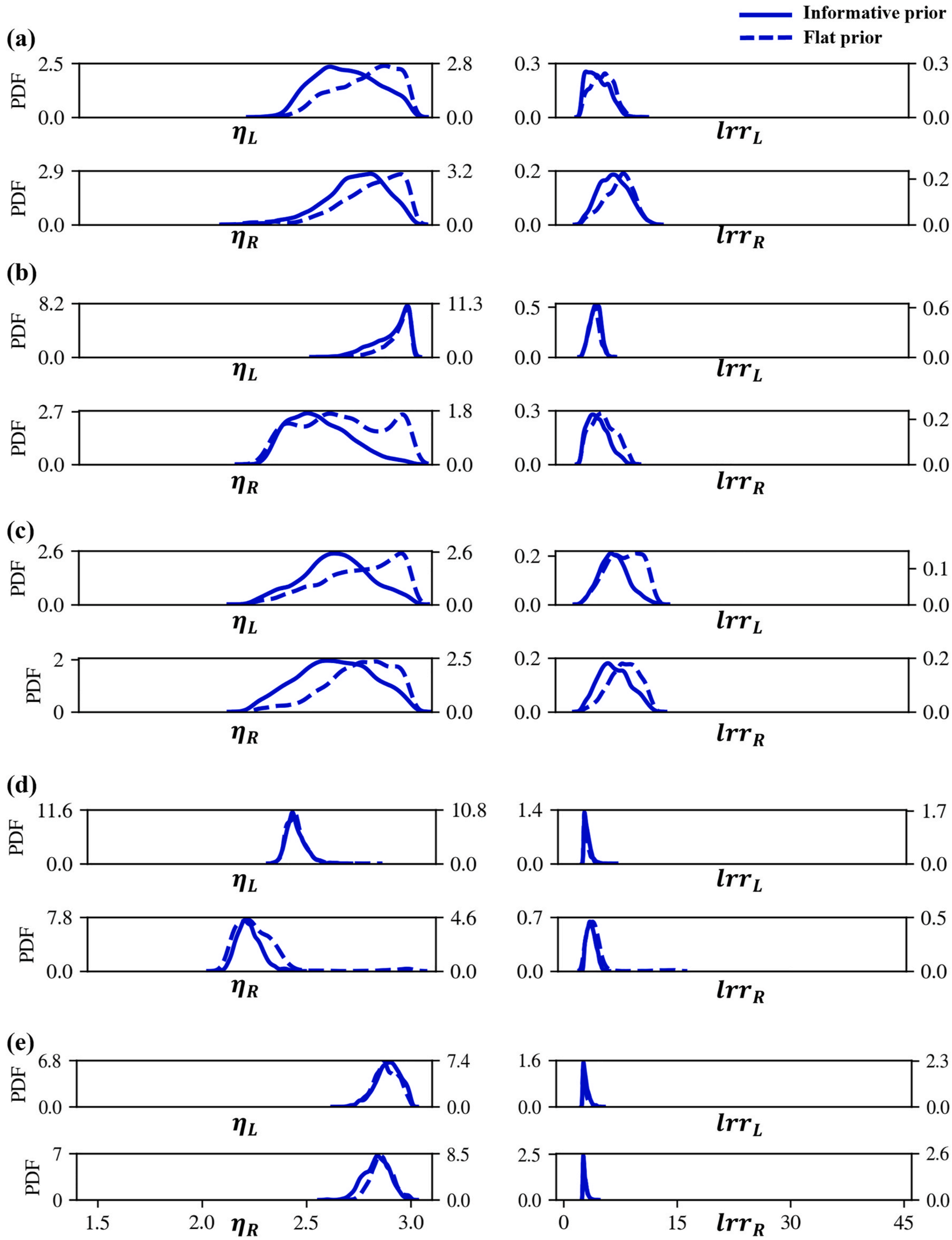

**Fig. 7.** Informative prior impact on the marginal posteriors in baseline. The left y-axis represents informative prior PDF, and the right y-axis represents flat prior PDF. Subplots are organized as follows: (a) D1, (b) D2, (c) D3, (d) D4, and (e) D5.

### 4.1. Gaussian process emulation

Statistical emulation modeling is a widely used approach to reduce complexity and enhance computational efficiency in mathematical models. The choice of an emulator depends on the characteristics of the dataset and the requirements of the mathematical model. For example, Paun et al. [36] conducted a comparative analysis of GPs and PCEs using the same 1D fluid dynamics model with two different terminal boundary conditions for both forward and inverse problems. They reported that GPs consistently performed slightly better than PCEs in every comparison. In this work, we used GPs due to its advantages over other methods and its relative flexibility in function approximation. GPs typically require smaller datasets for effective training; for instance, we used 3000 samples for training, whereas neural network architectures can require substantially more samples for the training process [37].

Emulating time-series data can require additional GP kernels to ac-count for both parameter and time-dependent correlations of the simulator [38]. Several studies have overcome this by combining GPs with PCA-based dimensionality reduction techniques ([39]; [40]). Previous studies [36,38,41] demonstrated that applying PCA to the output space does not have a significant effect on the accuracy of model predictions. Therefore, for our problem, we used PCA representations of our outputs to reduce the output dimensions from 140 to 20, thereby speeding up the training process. This also provides consistent theoretical links between the use of independent GPs with independent PCA components. We trained the GP model on area data but converted it to dynamic strain for parameter estimation, as strain is non-dimensional and could introduce identifiability issues during GP training. The GP accuracy for predicting test data varies among the subjects in baseline and CTEPH. As shown in Fig. 3(a and b) for pressure and area and Fig. 3 (c and d) for flows, the trends are similar among the subjects. In certain cases, some observations cause the outliers to reach a log(MSE) $= 10^{-1}$. However, on average, the log(MSE) remains relatively small among different cases. We note that some of the poor fits to test data are attributed to specific combinations of parameters that generate non-physiological signals in both magnitude and shape, such as unrealistic pressures (>120 mmHg) or large portions of negative flow in the proximal arteries. These results suggest that, in general, emulations align with the simulations across all four data sources, especially in the parameter domain near the inferred parameters.

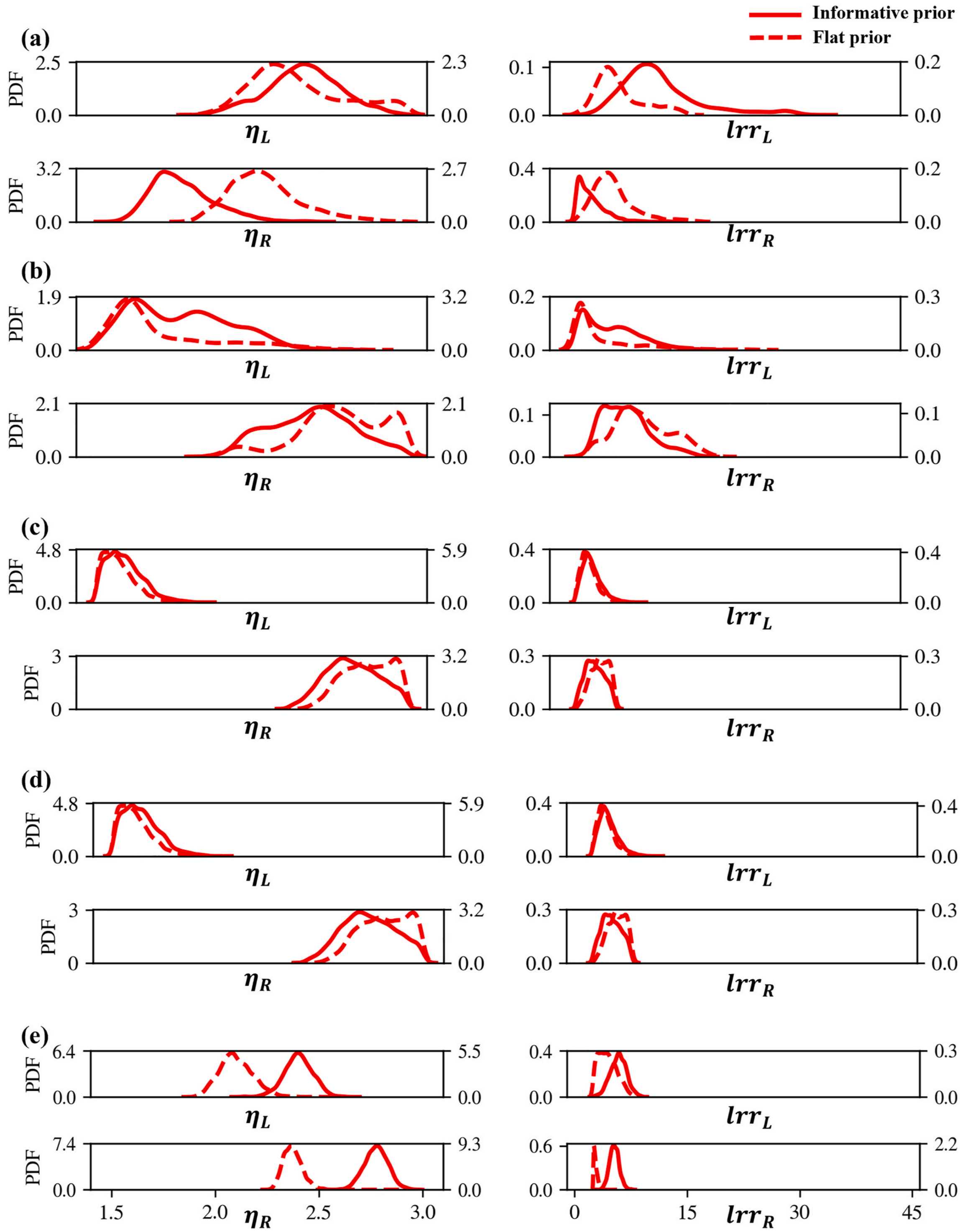

**Fig. 8.** Informative prior impact on the marginal posteriors in CTEPH. The left y-axis represents informative prior PDF, and the right y-axis represents flat prior PDF. Subplots are organized as follows: (a) D1, (b) D2, (c) D3, (d) D4, and (e) D5.

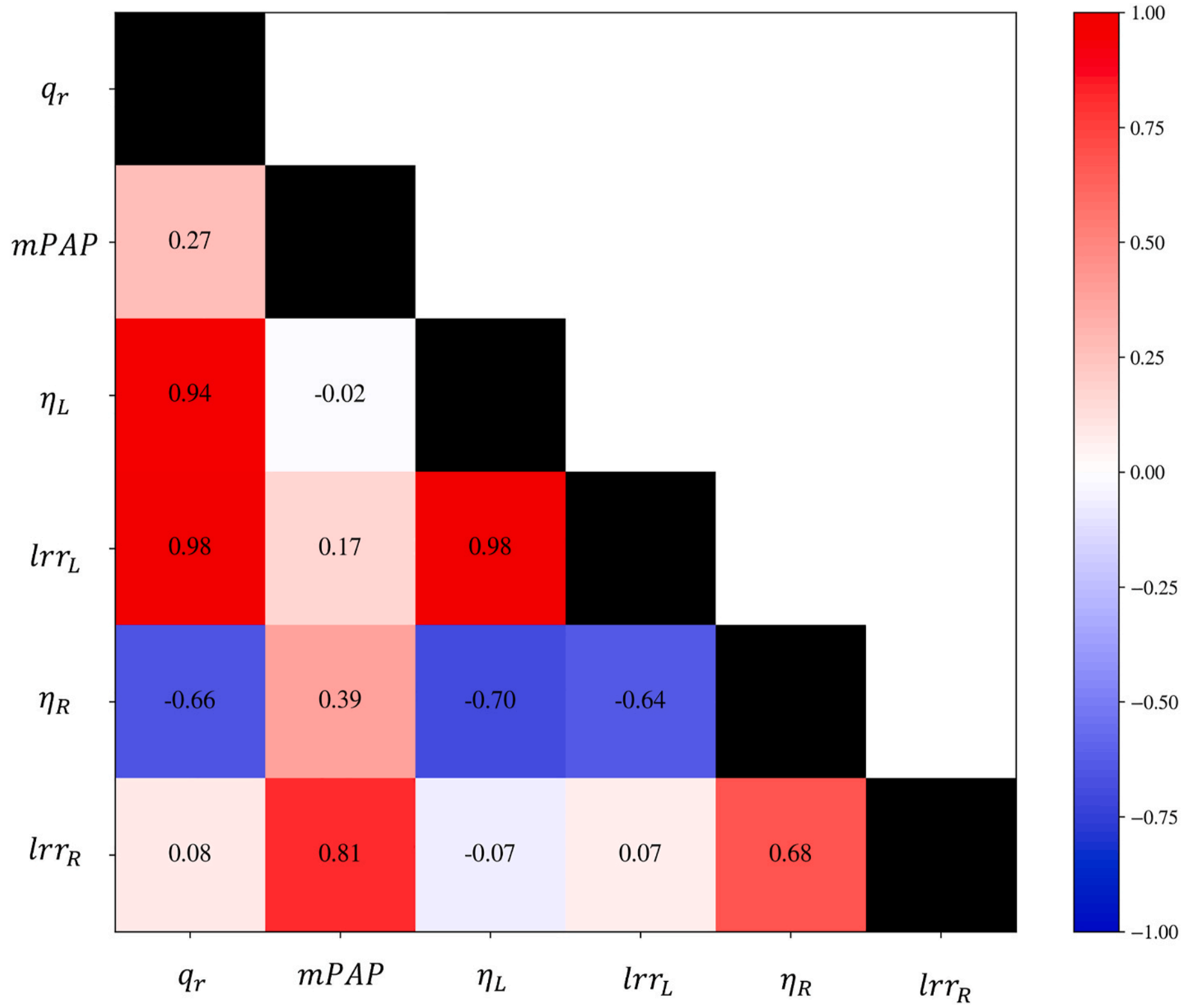

**Fig. 9.** Correlation analysis for the model outputs.

### 4.2. Parameter inference

Several studies have trained GP emulators to handle parameter estimation problems [42,43]. In the systemic circulation, several studies have used GPs to reduce computation time for quantifying uncertainties in their model predictions [44,45]. To date in the context of pulmonary blood flow simulation, studies have calibrated 1D hemodynamics models to normotensive and PH conditions [46–48], but only one prior study (by the authors, [1]) have calibrated a 1D model to CTEPH data. In addition, only a few studies [36,38] have calibrated parameters of the microvasculature, described by the structured tree, to hemodynamic data. In our study, we used a 1D fluid dynamics model with structured tree boundary conditions to model microvascular disease. Given the high computational cost of full Bayesian inversion with UQ, our study is among the few to combine GP emulation with multiple outputs (pressure, flow, and strain) from a PDE-based hemodynamics model. Furthermore, in comparison to other studies that used one set of parameters for the entire pulmonary circulation [25,46–48] we separated the left and right lung parameters to capture the heterogeneous nature of CTEPH.

We applied Bayesian inference with the emulator to infer the model parameters and quantify uncertainties in both parameters and outputs. This approach is particularly valuable because all clinical data, whether invasive or non-invasive, are inherently noisy due to limitations in measurement quality or methodology. Since measurement noise is rarely directly measurable or reported, we account for this uncertainty within the Bayesian inference process itself. We defined our likelihood functions such that each data source included a separate noise-variance term, instead of combining them into a single likelihood function with a common noise term [38,49]. Overall, the quality of fits was better with baseline data compared to CTEPH data, as shown in Figs. 4 and 5, respectively. One possible reason for this discrepancy is that although the dogs were of the same breed, age, and sex, they reacted differently during the development of CTEPH. For example, some cases required ~30,000 microspheres while others needed ~60,000 microspheres to reach CTEPH conditions [20].

Blood pressure is a critical factor, as PH is diagnosed based on elevated pressure levels. Consequently, most studies have calibrated their models using measured pressure signals [3,46,5]. In our study, the pressure measurements were limited to systolic and diastolic points, and we did not have access to the full pressure waveform. Nevertheless, we chose not to reduce the pressure output from the PDE model to these static points. Instead, we retained the full pressure signal to enable a richer uncertainty quantification analysis, such as generating credible intervals on the simulated outputs. In a recent study, Taylor-Lapole et al [2] employed a similar 1D fluid dynamics framework for the systemic circulation using structured tree boundary conditions. The patient data used for model calibration included systolic and diastolic pressure, along with four dynamic flow signals. Their model predictions revealed wider prediction intervals for pressure compared to flow. Similarly, in our study, we observed larger prediction intervals for pressure relative to other data sources, particularly in cases where the pressure fit was less accurate. Overall, almost all subjects showed reasonable agreement between the data and model predictions, including the prediction intervals, in the baseline condition. The exception was D2, which could be attributed to its unusually small, non-physiological pulse pressure (10 mmHg) compared to the other subjects.

As shown in Fig. 4, both LPA and RPA flow fits were reasonable in baseline, except for the LPA in D4. These discrepancies can be attributed to the significantly higher proportion of negative flow measured in the LPA for this subject compared to others. Fit qualities varied in CTEPH, even though the model captured the general rise and fall of the measured waveforms (Fig. 5). This discrepancy could be attributed to the impacts of CTEPH on the large vessels, since the large vasculature was kept unchanged from baseline to CTEPH. These effects were particularly evident in the LPA fits, which are likely attributed to worse remodeling in the left lung [20]. The quality of the strain fits follow a similar trend to the pressure fits. In the baseline condition, both pressure and strain fits are reasonable for four out of five subjects. Again, D2 shows the worst fit (Fig. 4), likely due to the very small pulse pressure. The combination of low pulse pressure, low strain, and relatively low stiffness is inconsistent with the reference area and inflow magnitude, which may explain why the emulator is unable to capture this set of measurement signals and requires further investigation. CTEPH results in Fig. 5 showed a similar trend in strain and pressure fits. Interestingly, despite the overestimation of pressure in D4, the emulated strain is nearly identical to the data.

### 4.3. CTEPH effects on the model parameters

We sought to assess the impact of CTEPH on microvascular parameters in the model. The microvasculature is believed to involve more complex mechanisms in PH progression compared to the large vessels [50,51]. Previous computational modeling studies [1,52,53,54] investigated CTEPH and hemodynamic changes in large arteries; however, none of them have explored microvascular disease in a patient-specific, CTEPH context.

The Murray exponent, $\eta$, specifies how the diameter varies across each bifurcation [22]. Theoretical analyses have demonstrated that the Murray exponent decreases curvilinearly with increasing Reynolds number (for inertial effects) and increasing Womersley number (for unsteady effects) [30,55]. We can interpret our results from Fig. 6 in the following context: in the left lung, we observed a significant reduction in the Murray exponent from baseline to CTEPH. Since we also observed a decrease in LPA flow, the Reynolds and Womersley numbers will only increase when there is a decrease in diameter. Hence, smaller Murray exponents under low flow conditions imply a reduction in small vessel diameter. This finding aligns with previous reports [20] highlighting more severe vascular obstruction in the left lung compared to the right lung and likely left-lung remodeling. The Murray exponent in the right lung exhibited a heterogeneous pattern, which may be attributed to less pronounced disease severity compared to the left lung.

As shown in Fig. 6, the $lrr_L$ value increased in all subjects from baseline to CTEPH, except D3, which may be attributed to only a slight pressure change from baseline to CTEPH in this subject. The observed increase in the length-to-radius ratio is likely due to elevated vascular resistance caused by a decrease in vessel radius relative to vessel length following the development of CTEPH. Consistent with the findings for the Murray exponent, this alteration was more pronounced in the left lung, further supporting the heterogeneous nature of disease severity between the two lungs. In addition, the $lrr$ value modulates the mean impedance, $Z(0)$, of the structured tree, thus reflecting increased resistance in the left lung due to CTEPH.

Variability in pulmonary blood flow serves as an early marker of lung disease or susceptibility to such conditions [56]. Yang et al [57] used 3D fluid dynamics simulation to evaluate the hemodynamic effects of pulmonary artery reconstruction in patients with Alagille syndrome and peripheral pulmonary stenosis, reporting that flow split in the pulmonary circulation can be an indicator of disease risk. In this study, we used the flow split index, $q_r$, as an indicator of microvascular disease severity. As illustrated in Fig. 9, $q_r$ exhibits a strong positive relationship with both left lung parameters. This relationship can be attributed to the impact of severe obstruction in the left lung, which significantly influenced both parameters. The lack of correlation between mPAP and $\eta_L$ provides with evidence that the radius exponent on the most affected side of the lung (left) does not linearly correlate with mPAP. However, we see a clear decrease in $\eta_L$ in subjects D2, D4, and D5 from baseline to CTEPH, as well as an increase in mPAP. This suggests that this factor does not contribute to mPAP increases linearly, but instead may be involved in nonlinear changes in mPAP. The positive correlation between mPAP and $lrr_R$ can be attributed to the remodeling that occurred in the right lung (the less obstructive lung), which experienced an increase in flow and to higher wall shear stress, which is suggested to play a role in vascular enlargement and elongation [58]. This remodeling increases impedance, requiring higher pressure to sustain flow. Furthermore, the strong positive correlation between $lrr_L$ and $\eta_L$, as well as $lrr_r$ and $\eta_r$, tell us that a proper calibration to hemodynamic data is linked to balancing between left and right lung parameters. This result is supported by Fig. 6, which shows that the posterior MAP values, and provides evidence that larger values of $lrr_r$ or $lrr_L$ correspond to larger MAP values in $\eta_r$ and $\eta_L$, respectively. These findings highlight the severity of disease in the left and right lungs, which may aid in surgical planning by informing decisions on left versus right lung treatment. Moreover, such frameworks can be extended to simulate treatment outcomes, including surgical interventions and vasodilator therapy, a common method of treatment in various PH pathologies. This can be combined with previous approaches that simulated clot removal in CTEPH [53]. When integrated with larger clinical datasets, they may help identify optimal treatment strategies for patients with CTEPH, guiding both surgical and pharmacological decision-making.

### 4.4. Effect of prior distribution

In Bayesian inference, an overly informative prior can disproportionately affect posterior estimates and subsequent conclusions about the parameters [59,60]. In contrast, the biggest benefit of Bayesian inference is the ability to use prior information. In our study, we use imaging data to inform the prior distributions for inference. We compared solutions to the inverse problem using uniform priors and data-informed priors as shown in Figs. 7 and 8 for both baseline and CTEPH, respectively. Overall, the data-informed prior had a greater influence on the CTEPH cases. We also compared the model outputs based on informative and flat priors for subjects that showed more noticeable changes in their marginal posteriors, D3 at baseline and D1 and D5 under CTEPH conditions. As shown in Supplementary Fig. S8, the output signals varied with the prior assumptions. Informative priors are beneficial and reflect a key strength of the Bayesian framework. It is important to note that our current dataset is small, and the informative prior was derived from this limited data. Expanding the dataset would enable the development of more accurate and generalizable informative priors. Furthermore, the use of informative priors transformed some marginal posterior distributions with long tails into more concentrated distributions, particularly in D2 and D5 for CTEPH. The sensitivity of left and right lung parameters to the prior distribution were divergent: the left lung parameters almost always shifted to the right, while right lung parameters exhibited variable behavior, including mostly leftward shifts (D1, D3, D4) and one rightward shift (D5).

## 5. Limitations

Our study has several limitations. First, time-varying pressure signals were not captured during data collection, and instead, only systolic and diastolic data was available. Calibrating the model using only systolic and diastolic pressures made the model calibration process more difficult and having a dynamic pressure signal would simplify the inverse problem and result in smaller prediction intervals for the pressure signal [61]. Second, we parameterized our model so that the entire left lung and right lung were described by lobe-specific parameters. Ideally, each vessel could have its own parameters, but this would make inference exceedingly difficult and likely introduce parameter identifiability issues [61]. Third, since our focus was on modeling microvascular disease in CTEPH, we fixed the geometry at baseline and inferred the parameters for both baseline and CTEPH. By doing so, we ignored alterations in the larger vessels after microsphere injection, which is not entirely realistic. Fourth, due to limited data, we applied a fixed set of assumptions, such as the area ratio and informative priors, derived from imaging data across all subjects. Incorporating subject-specific values for these parameters would enhance the study by making it more tailored to individual subjects, as each one ultimately has a unique vascular network. Finally, our sample size was small, making it difficult to draw solid conclusions, particularly in correlation analysis. A larger sample size would provide better insights and lead to a stronger understanding of how simulated quantities and model parameters correlate with disease severity.

## 6. Conclusion

We developed a rapid Bayesian framework that combines emulation with a 1D fluid dynamics model of the pulmonary circulation to efficiently estimate microvascular parameters and their associated uncertainties in CTEPH. Microvascular disease in CTEPH is characterized by narrowed vessel diameters and high stiffness, which persist even after surgical removal of pulmonary emboli in the larger vessels. Our approach explicitly estimates two parameters, $\eta$ and $lrr$, which describe the geometry and branching patterns of the microvasculature. We successfully calibrated the model against measured data for both baseline and CTEPH cases with reasonable accuracy. The calibrated values of these parameters consistently shift from baseline to CTEPH, indicating microvascular narrowing and elevating impedance. Moreover, our findings suggest that the severity of CTEPH, as indicated by changes in mPAP and $q_r$, is directly linked to these parameter changes. In this dataset, we show that microvascular parameters not only track the transition from baseline to CTEPH but also correlate with disease severity, linking microvascular structure to the functional capacity of the pulmonary circuit. By understanding the correlation between measurable data and microvascular narrowing parameters, patient-specific assessments can be informed, the effects of potential therapies can be simulated and evaluated, and a crucial step can be taken toward unraveling the complex mechanisms of microvascular disease in CTEPH. This lays the foundation for future synergistic studies aimed at developing precise computational diagnostic tools and targeted interventions tailored to individuals with CTEPH.

## CRediT authorship contribution statement

**Amirreza Kachabi:** Writing – original draft, Visualization, Validation, Software, Methodology, Investigation, Formal analysis, Data curation, Conceptualization. **Sofia Altieri Correa:** Software, Data curation. **Naomi C. Chesler:** Writing – review & editing, Supervision, Resources, Funding acquisition. **Mitchel J. Colebank:** Writing – review & editing, Supervision, Software, Methodology, Conceptualization.

## Ethical Statement for Solid State Ionics

1) This material is the authors' own original work, which has not been previously published elsewhere.
2) The paper is not currently being considered for publication elsewhere.
3) The paper reflects the authors' own research and analysis in a truthful and complete manner.
4) The paper properly credits the meaningful contributions of co-authors and co-researchers.
5) The results are appropriately placed in the context of prior and existing research.
6) All sources used are properly disclosed (correct citation). Literally copying of text must be indicated as such by using quotation marks and giving proper reference.
7) All authors have been personally and actively involved in substantial work leading to the paper, and will take public responsibility for its content.

The violation of the Ethical Statement rules may result in severe consequences.

To verify originality, your article may be checked by the originality detection software iThenticate. See also http://www.elsevier.com/editors/plagdetect.

I agree with the above statements and declare that this submission follows the policies of Solid State Ionics as outlined in the Guide for Authors and in the Ethical Statement.

## Declaration of competing interest

The authors declare that they have no known competing financial interests or personal relationships that could have appeared to influence the work reported in this paper.

## Acknowledgements

This work was funded by the National Institutes of Health NIBIB grants R01HL154624 (NCC) and R01 HL147590 (NCC). M.J.C. was supported in part by TL1 TR001415 through the National Center for Research Resources and the National Center for Advancing Translational Sciences, National Institutes of Health. The content is solely the responsibility of the authors and does not necessarily represent the official views of the NIH.

## Appendix A. Supplementary data

Supplementary data to this article can be found online at https://doi.org/10.1016/j.compbiomed.2025.110552.